\documentclass[pre,twocolumn,showpacs,superscriptaddress,10pt]{revtex4-1}
\usepackage[colorlinks,citecolor=blue,linkcolor=red,dvipdfm]{hyperref}
\usepackage{amsmath,amssymb,graphicx}
\usepackage{times}

\begin{document}

\title{Localized dark solitons and vortices in defocusing media with
spatially inhomogeneous nonlinearity}
\author{Jianhua Zeng}
\email{\underline{zengjh@opt.ac.cn}}
\affiliation{State Key Laboratory of Transient Optics and Photonics, Xi'an
Institute of Optics and Precision Mechanics of CAS, Xi'an 710119, China\\
}

\author{ Boris A. Malomed}
\email{\underline{malomed@post.tau.ac.il}}
\affiliation{Department of Physical Electronics, School of Electrical
Engineering, Faculty of Engineering, Tel Aviv University, Tel Aviv 69978,
Israel\\
Laboratory of Nonlinear-Optical Informatics, ITMO University, St. Petersburg 197101,
Russia}

\begin{abstract}
Recent studies have demonstrated that defocusing cubic nonlinearity with
local strength growing from the center to the periphery faster than $r^{D}$,
in space of dimension $D$ with radial coordinate $r$, supports a vast
variety of robust bright solitons. In the framework of the same model, but
with a weaker spatial-growth rate, $\sim r^{\alpha }$ with $\alpha \leq D$,
we here test the possibility to create stable\textit{\ localized continuous
waves} (LCWs) in one- and two-dimensional (1D and 2D) geometries, \textit{%
localized dark solitons} (LDSs) in 1D, and \textit{localized dark vortices}
(LDVs) in 2D, which are all realized as loosely confined states with a
divergent norm. Asymptotic tails of the solutions, which determine the
divergence of the norm, are constructed in a universal analytical form by
means of the Thomas-Fermi approximation (TFA). Global approximations for the LCWs,
LDSs, and LDVs are constructed on the basis of interpolations between
analytical approximations available far from (TFA) and close to the center.
In particular, the interpolations for the 1D LDS, as well as for the 2D
LDVs, are based on a \textquotedblleft deformed-tanh" expression, which is
suggested by the usual 1D dark-soliton solution. The analytical
interpolations produce very accurate results, in comparison with numerical
findings, for the 1D and 2D LCWs, 1D LDSs, and 2D LDVs with vorticity $S=1$.
In addition to the 1D fundamental LDSs with the single notch, and 2D
vortices with $S=1$, higher-order LDSs with multiple notches are found too,
as well as double LDVs, with $S=2$. Stability regions for the modes under
the consideration are identified by means of systematic simulations, the
LCWs being completely stable in 1D and 2D, as they are ground states in the
corresponding settings. Basic evolution scenarios are identified for those
vortices which are unstable. The settings considered in this work may be
implemented in nonlinear optics and in Bose-Einstein condensates.
\end{abstract}

\maketitle


\section{Introduction}

It is commonly believed that the formation of localized waves (alias
solitons) in uniform media is a result of the balance between diffraction
and self-focusing or defocusing nonlinearity \cite{FPC}. Following this
concept, it has been established that defocusing nonlinearity creates dark
solitons, while the self-focusing nonlinearity is necessary for the
existence of bright solitons in homogeneous media.

The situation may be different in inhomogeneous media. First, inhomogeneity
can be represented by spatially periodic linear potentials, induced by
photonic crystals in optics \cite{PC}, by gratings built into plasmonic
waveguides for surface plasmon waves \cite{plasmon}, and by optical lattices
in atomic Bose--Einstein condensates (BECs) \cite{BEC, BEC-Rev1, BEC-Rev2}
or Fermi gases \cite{Fermi}. Owing to the straightforward realization in a
vast variety of physical systems, periodic potentials play an increasingly
important role in manipulations of different kinds of waves and solitons
\cite{G-Assanto,Peli}. It has been demonstrated that the periodic
potentials, in a combination with the self-focusing or defocusing material
nonlinearity, give rise to various species of bright solitons \cite%
{BEC-Rev1, BEC-Rev2, soliton-Rev1, soliton-Rev2, soliton1, soliton2},
including ordinary ones, residing in the semi-infinite gap of the system's
linear spectrum, and gap solitons in finite bandgaps. In particular, the
formation of bright gap solitons in the system with the defocusing sign of
the nonlinearity may be interpreted as a result of the reversal of the sign
of the effective dispersion under the action of the periodic potential \cite%
{uniform-GS}-\cite{Salerno}.

Another possibility for the creation of bright solitons is offered by
nonlinear lattices (see a review \cite{NL} and references therein). These
are nonlinear counterparts of linear periodic potentials, which are induced
by spatially periodic modulations of the local strength and, possibly, sign
of the nonlinearity. In optics, nonlinear lattices may be engineered by
means of properly designed photonic-crystal structures (e.g., filling voids
in photonic crystals by solid \cite{Russell} or liquid \cite{phot-cryst}
materials with different values of the Kerr coefficient, or the coefficient
accounting for the quadratic nonlinearity \cite{chi^2}). Another possibility
is to use inhomogeneous distributions of nonlinearity-enhancing dopants \cite%
{dopant} (note, in particular, that appropriate dopants may induced
defocusing on top of a self-focusing background \cite{Gaetano}). In BEC,
similar nonlinearity landscapes can be induced by the Feshbach resonance in
spatially nonuniform optical \cite{nonuniform-Feshbach} or magnetic \cite%
{Feshbach-magnetic} fields. In the framework of one-dimensional (1D)
settings, nonlinear lattices support bright solitons under various
conditions \cite{soliton-NL}.

On the contrary to the commonly adopted principle that pure defocusing
nonlinearities, without the help of linear potentials, cannot produce bright
solitons, it was demonstrated that media with a pure self-repulsive
(defocusing) spatially inhomogeneous nonlinearity, whose local strength
grows from the center to the periphery at any rate faster than $r^{D}$ ($r$
is the radial coordinate), can support a great variety of robust
self-trapped modes in the space of dimension $D$, including 1D fundamental
and higher-order (dipole and multipole) solitons, 2D solitary vortices with
arbitrarily high topological charges \cite{soliton-Defo1}, as well as
sophisticated 3D modes, such as soliton gyroscopes \cite{Defo9} and
skyrmions, i.e., vortex rings with intrinsic twist \cite{Defo8}. Such modes
exist due to the balance between the spatially inhomogeneous repulsive
nonlinearity and linear dispersion/diffraction. A characteristic feature of
these self-trapped modes is \textit{nonlinearizability} of the underlying
equations for their decaying tails, in contrast to the usual bright solitons
in media with uniform or periodically modulated nonlinearities, where
solitons are restricted to the semi-infinite or finite bandgaps of the
corresponding linear spectrum \cite{BEC-Rev1, BEC-Rev2, soliton-Rev1,
soliton-Rev2, soliton1, soliton2}. In fact, the nonlinearizability of the
tails makes the concept of the linear spectrum irrelevant for the solitons
supported by repulsive nonlinearities with the spatially growing local
strength.

The exploration of bright solitons and solitary vortices supported by such a
scheme has been extended to a variety of physical settings, see Refs. \cite%
{soliton-Defo1}-\cite{Defo13} and references therein. However, all the
studies in this field were, thus far, restricted to the above-mentioned
condition of the steep spatial modulation of the self-defocusing, whose
local strength must grow with $r$ faster than $r^{D}$. Actually, this
condition secures the convergence of the soliton's total norm for bright
solitons. Weakly localized states, possible in the presence of a more gentle
modulation, $\sim r^{\alpha }$ with $\alpha \leq D$, have not been studied
yet. Because their total norm diverges (they are localized too loosely for
the convergence of the norm), like in the usual continuous-wave states and
dark solitons \cite{dark}, such spatially even (symmetric) and odd
(antisymmetric) modes may be considered, respectively, as \textit{localized
continuous-wave} (LCW) states, and as \textit{localized dark solitons}
(LDSs). In particular, the LCW configurations represent the ground state in
the present settings. The consideration of this system is relevant, as it
may be easier to implement a gentle modulation profile in the experiment
than its above-mentioned steep counterpart, and the LDS offers an essential
extension of the well-elaborated concept of dark solitons in the uniform 1D
space \cite{dark}. This is the subject of the present work. We here focus on
1D and 2D media with the defocusing cubic nonlinearity subject to the
moderate spatial modulation. In these settings, 1D LDS solutions and 2D
vortex modes (\textit{localized dark vortices}, LDVs) with vorticity $S=1$
can be obtained in an approximate analytical form, which is validated by
comparison with numerical results. More complex states, such as
higher-order1D solitons and vortices with $S>1$, are constructed numerically.

The paper is organized as follows. In Sec. II, we introduce the model and
report approximate analytical solutions. Tails of the weakly localized
states are produced in a universal form, which does not depend on the
spatial dimension, nor on the parity (spatial symmetry or antisymmetry) of
the underlying solution, by the Thomas-Fermi approximation (TFA). For the
global structure of the LCWs, LDSs, and LDVs we develop approximations based
on interpolations between asymptotic forms available close to the center and
far from it. In particular, a \textquotedblleft deformed-tanh" interpolation
produces quite accurate results for 1D LDS and 2D LDV\ with $S=1$. In Sec.
III, numerical results are reported for the shape and stability of 1D
fundamental and higher-order LCWs and LDSs. Numerical findings for 2D states
are presented in Sec. IV. The paper is concluded by Sec. V.

\section{The model and analytical approximations}

\subsection{The underlying equations}

The model is based on the generalized nonlinear Schr\"{o}dinger (NLS) or
Gross-Pitaevskii equation for the mean-field wave function $\psi (\mathbf{r}%
,z)$, written in the scaled form:
\begin{equation}
i\psi _{z}=-\frac{1}{2}\nabla ^{2}\psi +g(r\mathbf{)}|\psi |^{2}\psi .
\label{GPE}
\end{equation}%
The mode is cast in the optics notation, with the evolution variable, $z$,
realized as the propagation distance (for matter waves, $z$ is replaced by
time, $t$), $g(r\mathbf{)}>0$ being the local strength of the defocusing
cubic term, which is specified below, and Laplacian $\nabla ^{2}=\partial
_{x}^{2}+\partial _{y}^{2}$ acts on transverse coordinates $\left\{
x,y\right\} $ in the bulk medium, with $r=\sqrt{x^{2}+y^{2}}$. Stationary
solutions to Eq. (\ref{GPE}) with real propagation constant $k$ are sought
for as $\psi (x,y,t)=\phi (x,y)\exp (ikz)$, where the stationary wave
function (generally speaking, it may be complex) obeys its own equation:
\begin{equation}
k\phi =\frac{1}{2}\nabla ^{2}\phi -g(r)|\phi |^{2}\phi .  \label{SE}
\end{equation}%
Equation (\ref{SE}) can be derived from the respective Lagrangian,
\begin{equation}
L=\int \int \left[ k|\phi |^{2}+\frac{1}{2}\left\vert \nabla \phi
\right\vert ^{2}-\frac{1}{2}g(r)|\phi |^{4}\right] dxdy.  \label{L}
\end{equation}%
The 1D model corresponds to an obvious one-dimensional reduction of Eqs. (%
\ref{GPE})-(\ref{L})

To illustrate the concept of the loosely localized solitons, we note that
the 2D version of Eq. (\ref{SE}) with%
\begin{equation}
g(r)=g_{0}r^{2},  \label{2}
\end{equation}%
which corresponds to the critical case of $\alpha =D$, gives rise to a
family of weakly singular \emph{exact} LDV solutions with vorticity $S=\pm 1$%
:%
\begin{equation}
\phi \left( r,\theta \right) =\sqrt{-k/g_{0}}r^{-1}e^{\pm i\theta },
\label{exact}
\end{equation}%
where $\theta $ is the angular coordinate, and wavenumber $k$ may take any
\emph{negative} value.

\subsection{The Thomas-Fermi approximation}

In the framework of the TFA, which neglects derivatives in stationary
equation (\ref{SE}), it gives rise to real self-trapped solutions in the
form of
\begin{equation}
\phi _{\mathrm{TFA}}(r)\approx \sqrt{-k/g(r)},  \label{TFA}
\end{equation}%
which is relevant at large $r$, where it is asymptotically exact, i.e.,
\begin{equation}
\left[ \phi (r)-\phi _{\mathrm{TFA}}(r)\right] /\phi _{\mathrm{TFA}%
}(r)\rightarrow 0~~\mathrm{at~~}r\rightarrow \infty   \label{r-->infty}
\end{equation}%
\cite{soliton-Defo1}. Because the defocusing nonlinearity implies $g(r)>0$,
Eq. (\ref{TFA}) holds for $k<0$, cf. exact vortex solution (\ref{exact}).
The TFA produces universal results for the tails, which, as mentioned above,
do not depend on the spatial dimension, nor on the global structure of the
underlying solution [e.g., whether it is an LCW, or a fundamental LDS, or
its higher-order counterpart (see below), or an LDV].

For modulation profiles $g(r)$ which have a power-law asymptotic form at $%
r\rightarrow \infty $,%
\begin{equation}
g(r)=g_{0}r^{\alpha },  \label{asympt}
\end{equation}%
with $\alpha \leq D$ ($r$ is realized as $|x|$ for $D=1$), Eq. (\ref{TFA})
gives rise to slowly decaying tails in the form of%
\begin{equation}
\phi (r)\approx \sqrt{\frac{-k}{g_{0}}}r^{-\alpha /2}\left[ 1-\frac{\alpha }{%
16}\left( \alpha +2\left( 2-D\right) \right) r^{-2}\right] ,  \label{postTFA}
\end{equation}%
where the second term in the parentheses represents the first post-TFA
correction, that corroborates relation (\ref{r-->infty}). It follows from
Eq. (\ref{postTFA}) that the critical point, $\alpha =D$, which separates
states with convergent and divergent values of the norm (or integral power,
in terms of optics), $N=\int \phi ^{2}\left( \mathbf{r}\right) d\mathbf{r}$,
is an \emph{exact} one (the correction to the TFA does not affect this
point).

The divergence of $N$ for $\alpha <D$ can be estimated by introducing an
overall size, $R$, of a \textit{truncated} version of the system,\ with $%
r\leq R$ (obviously, any physical system has a finite size). Then, Eq. (\ref%
{TFA}) yields%
\begin{equation}
N_{D}(\alpha )\approx \frac{2\pi ^{D-1}|k|}{g_{0}\left( D-\alpha \right) }%
R^{D-\alpha },  \label{ND}
\end{equation}%
for $D=1$ or $2$ ($|k|$ is written here, as $k$ is negative), which
explicitly displays the divergence at $R\rightarrow \infty $. In the limit
case of $\alpha =D$, Eq. (\ref{ND}) is replaced by%
\begin{equation}
N_{\alpha =D}\approx \frac{2\pi ^{D-1}|k|}{g_{0}}\ln \left( R/r_{0}\right) ,
\label{ln}
\end{equation}%
where $r_{0}$ is an internal size of the mode; for instance, it is $r_{0}=S/%
\sqrt{2|k|}$ for vortex (\ref{TFA-2D}), see below. We note that, while the
truncation, of course, determines the $N(\alpha )$ dependence in the case of
$\alpha \leq D$, it does not strongly affect the shape of the modes under
the consideration, if $R$ is large enough, as the shapes of the modes decays
at large $r$ anyway, even if relatively slowly. This conclusion is
corroborated, in particular, by the shapes displayed below in Figs. \ref%
{Fig1}, \ref{Fig3}, \ref{Fig6} and \ref{Fig8}.

It is worthy to note that expressions (\ref{ND})\ and (\ref{ln}) satisfy the
\textit{anti-Vakhitov-Kolokolov} (anti-VK) condition, $dN/dk<0$, which is a
necessary stability condition for localized modes supported by defocusing
nonlinearities \cite{antiVK}. In fact, the anti-VK condition is sufficient
for the stability of ground states, but it may not be sufficient for excited
states, see below.

In the 2D version of the model, LDVs states with integer vorticity $S$ are
looked for as%
\begin{equation}
\psi \left( r,\theta ,z\right) =e^{ikz+iS\theta }\varphi _{S}(r),  \label{S}
\end{equation}%
with real amplitude $\varphi (r)$ obeying the following equation:
\begin{equation}
-k\varphi _{S}=-\frac{1}{2}\left( \frac{d^{2}\varphi _{S}}{dr^{2}}+\frac{1}{r%
}\frac{d\varphi _{S}}{dr}-\frac{S^{2}}{r^{2}}\varphi _{S}\right)
+g(r)\varphi _{S}^{3}\ .  \label{vort}
\end{equation}%
For $g(r)$ taken as per Eq. (\ref{2}), and $S^{2}=1$, an exact solution of
Eq. (\ref{vort}) is given by Eq. (\ref{exact}). In the general case, a
solution to Eq. (\ref{vort}) can again be looked for by means of the TFA,
neglecting the radial derivatives:%
\begin{equation}
\varphi _{S}^{2}(r)=\left\{
\begin{array}{c}
-\left[ 1/g(r)\right] \left[ k+S^{2}/\left( 2r^{2}\right) \right] ,~~\mathrm{%
at}~~r^{2}>-S^{2}/\left( 2k\right) , \\
0,~~\mathrm{at}~~r^{2}<-S^{2}/\left( 2k\right) ,%
\end{array}%
\right.   \label{TFA-2D}
\end{equation}%
cf. the application of the TFA to vortex solutions in Refs. \cite{TFA,Defo9}.

\subsection{Global interpolations}

The TFA is relevant for the outer zone, where the derivatives (diffraction,
in the optics model, or kinetic energy, in the BEC system) may be neglected.
As an attempt to construct a global approximation, we will try an
interpolation which goes over into the TFA at large $r$, and matches a
correct analytical form of the solution in its inner zone. The interpolation
for the LCW (the ground state), corresponding to the modulation profile (\ref%
{asympt}), is based on the following simplest expression, which is
compatible with the TFA asymptotic form (\ref{TFA}) and the fact that the
LCW state must be free of singularities and feature a maximum at $r=0$:
\begin{equation}
\phi (r)=\sqrt{\frac{|k|}{g_{0}}}\left( r^{2}+r_{0}^{2}\right) ^{-\alpha /4}.
\label{LCW}
\end{equation}%
This expression does feature spatial dimension $D$ (recall that $r^{2}$ is
replaced by $x^{2}$ in 1D). However, the value of constant $r_{0}$, which is
determined by the substitution of this expression in Eq. (\ref{SE}) at $r=0$
(or $x=0$, in the 1D case), depends on the $D$:%
\begin{equation}
r_{0}^{2}=D\alpha /\left( 4|k|\right) .  \label{r0}
\end{equation}%
In fact, interpolation (\ref{LCW}) may be used for the bright solitons at $%
\alpha >D$ too, although in that case it turns out to be less accurate, see
below.

It is relevant to calculate the total norm of the truncated version ($r\leq R
$) of expression (\ref{LCW}) with $\alpha \leq 2$ for $D=2$ (for $D=1$ the
analytical result is too cumbersome):%
\begin{gather}
N_{D=2}(\alpha <2)  \notag \\
=\frac{2\pi |k|}{g_{0}\left( 2-\alpha \right) }\left[ \left( R^{2}+\frac{%
\alpha }{2|k|}\right) ^{1-\alpha /2}-\left( \frac{\alpha }{2|k|}\right)
^{1-\alpha /2}\right] ,  \label{Nans1} \\
N_{D=2}(\alpha =2)=\frac{\pi |k|}{g_{0}}\ln \left( \frac{2|k|}{\alpha }%
R^{2}+1\right) .  \label{Nans2}
\end{gather}%
Obviously, in the limit of $\left( 2|k|/\alpha \right) R^{2}\gg 1$,
expressions (\ref{Nans1}) and (\ref{Nans2}) carry over into Eqs. (\ref{ND})
and (\ref{ln}), respectively, produced by the TFA.

For the 2D LDV with vorticity $S\geq 1$, as well as for the 1D LDS, a
natural form of the interpolation, that we call \textquotedblleft deformed
tanh", is suggested by the hyperbolic-tangent solution for the usual dark
soliton in the 1D uniform space:
\begin{equation}
\left( \varphi _{S}(r)\right) _{\mathrm{tanh}}=\sqrt{\frac{|k|}{g(r)}}\left[
\tanh (\lambda r)\right] ^{S},  \label{tanh}
\end{equation}%
where $1/\lambda $ determines the radius of the LDS core, which is used
below as a fitting parameter, while comparing Eq. (\ref{tanh}) to numerical
results. In the 1D geometry, $\left( \varphi _{S}(r)\right) _{\mathrm{tanh}}$
in Eq. (\ref{tanh}) is replaced by
\begin{equation}
\phi _{\mathrm{tanh}}(x)=\sqrt{\frac{|k|}{g(x)}}\tanh (\lambda x),
\label{tanh-1D}
\end{equation}%
while $r$ is replaced by $x$ as the argument of $g(r)$ and $\tanh (\lambda r)
$.
\begin{figure}[tbp]
\begin{center}
\includegraphics[width=1.04\columnwidth]{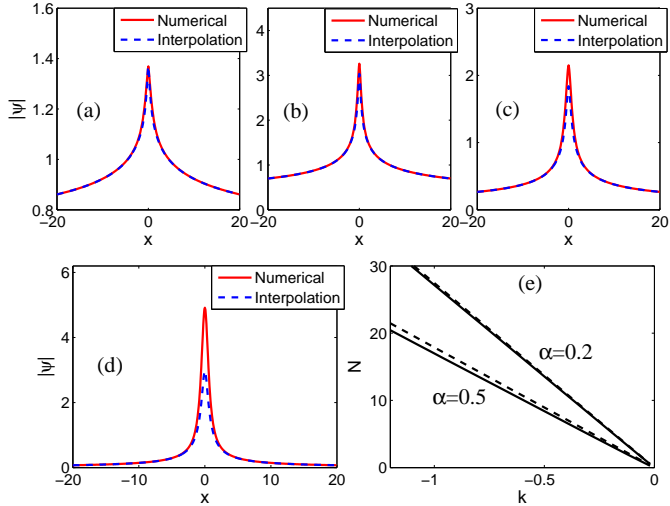}
\end{center}
\caption{(Color online) The comparison of the analytical interpolation for
1D localized-continuous-wave (LCW) states, given by Eq. (\protect\ref{LCW}),
with their numerically found counterparts, for (a) $\protect\alpha =0.2$, $%
k=-1.35$, (b) $\protect\alpha =0.65$, $k=-3.4$, (c) $\protect\alpha =1$, $%
k=-2.1$, and (d) $\protect\alpha =2$, $k=-2.1$ (the latter case, with $%
\protect\alpha >D=1$, corresponds to a bright soliton). (e) The numerically
found dependence $N(k)$ for the 1D LCW at different values of $\protect%
\alpha $, with norm $N$ computed in a finite domain, $|x|\leq 20$. The
dashed lines show the analytical counterpart, given by Eq. (\protect\ref{ND}%
) with $D=1$ and $R=20$. This figure and all others pertain to $g_{0}=1$ in
Eq. (\protect\ref{asympt}), fixed by scaling.}
\label{Fig1}
\end{figure}

\begin{figure}[tbp]
\begin{center}
\includegraphics[width=1\columnwidth]{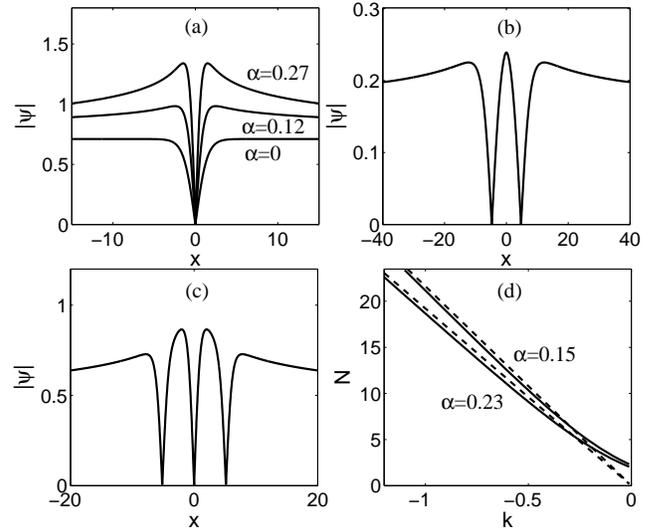}
\end{center}
\caption{(a) Shapes of 1D fundamental (single-notch) localized dark solitons
(LDSs) with different values of parameters ($\protect\alpha =0$, $k=-0.5$, $%
\protect\alpha =0.12$, $k=-1.1$, and $\protect\alpha =0.27$, $k=-2.1$). The
comparison of these numerically found profiles with their approximate
analytical counterparts is displayed in Fig. \protect\ref{Fig3}. Panel (a)
includes, for the sake of comparison, the commonly known exact dark soliton
for $\protect\alpha =0$. Panels (b,c) present examples of higher-order LDSs
with different numbers of notches: (b) two notches for $\protect\alpha =0.25$%
, $k=-2.2$, and (c) three notches for $\protect\alpha =0.3$, $k=-1$. (d) The
norm ($N$) vs. the propagation constant ($k$) for 1D fundamental LDSs in the
truncated system with different values of $\protect\alpha $. Dashed straight
lines show the respective analytical approximations given by Eq. (\protect
\ref{ND}).}
\label{Fig2}
\end{figure}

Below, results produced by numerical computations are displayed and compared
to these analytical approximations. We stress that the validity of the
numerical schemes needs to be carefully checked in the present setting, as
properly handling boundary conditions (b.c.) for weakly localized modes with
slowly decaying tails is a known challenging problem in simulations of
nonlinear partial differential equations, such as the NLS equation (\ref{GPE}%
). In particular, we have inferred that the usual split-step fast Fourier
transform method with periodic b.c. does not apply to the present model, and
the finite-difference method with Neumann or Dirichlet b.c. is not an
appropriate one either. To resolve the issue, we have developed a
finite-difference method with dynamical b.c. for robust simulations of Eq. (%
\ref{GPE}). These b.c. are defined as $\frac{\partial }{\partial z}\left(
\frac{\partial \psi }{\partial r}|_{\mathsf{BP}}\right) =\frac{\partial }{%
\partial z}\left( \frac{\partial \psi }{\partial r}|_{\mathsf{BP}-1}\right) $%
, where subscripts $\mathsf{BP}$ and $\mathsf{BP}-1$ pertain to the boundary
point and the inner one adjacent to it (more technical details will be
presented elsewhere). It is relevant to mention that our numerical codes correctly
reproduce 1D and 2D bright solitons (with finite norms) for $\alpha >$ $D$, which
were found in earlier works \cite{soliton-Defo1}, \cite{Defo5}-\cite{Defo10}.
\begin{figure}[tbp]
\begin{center}
\includegraphics[width=1\columnwidth]{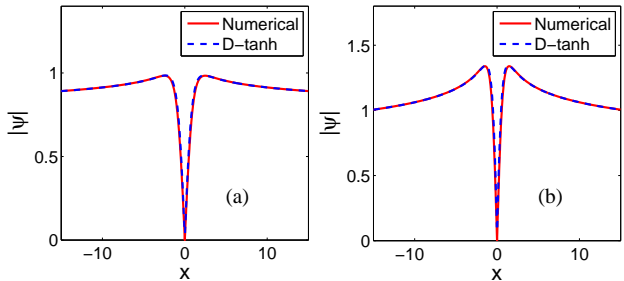}
\end{center}
\caption{(Color online) The comparison of numerically computed solutions for
the 1D fundamental localized dark solitons with the analytical prediction
based on the deformed-tanh interpolation (labeled \textquotedblleft D-tanh"
in the figure), given by Eq. (\protect\ref{tanh}) for (a) $\protect\alpha %
=0.12$ and $k=-1.1$, with $\protect\lambda =1.1$, and (b) $\protect\alpha %
=0.27$ and $k=-2.1$, with $\protect\lambda =1.5$. }
\label{Fig3}
\end{figure}
\begin{figure}[tbp]
\begin{center}
\includegraphics[width=1\columnwidth]{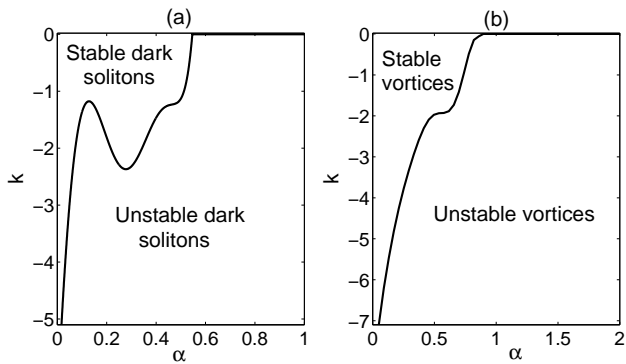}
\end{center}
\caption{ (a) The stability border for the 1D fundamental LDSs. (b) The same
for the 2D localized dark vortices (LDVs) with charge $S=1$. The dark
solitons and vortices are stable above the respective borders. While the
shape of the stability borders is relatively complex, the results meet the
natural condition that all the modes are stable in the limit of $\protect%
\alpha \rightarrow 0$, in compliance with the commonly known fact that dark
1D\ solitons and 2D vortices are completely stable in the case of the
spatially uniform self-defocusing. }
\label{Fig4}
\end{figure}

\section{Numerical results for one-dimensional localized modes}

In this section we report numerical results for the 1D model with modulation
profile (\ref{asympt}), in which $g_{0}=1$ is fixed by means of rescaling.
The LCW and LDS states (both fundamental and higher-order ones, in the
latter case) were produced as numerical solutions of stationary equation (%
\ref{SE}). The same stationary solutions can also be found by means of
integrating NLS equation (\ref{GPE}) in imaginary time, fixing propagation
constant $k$, which is another well-known numerical method for producing
stationary solutions \cite{IT}. The stability of the so found solutions was
then checked by means of systematic simulations of Eq. (\ref{GPE}) for the
evolution of perturbed solutions in real time, using the above-mentioned
finite-difference numerical scheme with the dynamical b.c. The spatial and
time steps were taken as $\Delta x=\Delta y=0.08$ and $\Delta z=0.001$,
using different sizes of the integration domain, which are appropriate in
different cases, as specified below.

Numerical simulations reveal that the shape of 1D LCW states conspicuously
change with the increase of the nonlinearity-modulation power, $\alpha $,
featuring growth of the amplitude and decrease of the width. Naturally, the
LCW states carries over into a fundamental bright soliton, with the
convergent norm, at $\alpha >D$. These features are clearly seen in Figs. %
\ref{Fig1}(a-d). Comparison between the numerical solutions for the 1D LCW
states and the analytical approximation based on Eqs. (\ref{LCW}) and (\ref%
{r0}) is shown in Figs. \ref{Fig1} too. It is seen that the LCW profiles
predicted by the analytical interpolation provide a good fit to their
numerical counterparts for all $\alpha \leq 1$, and the corresponding
expression (\ref{ND}) for the norm, calculated in the truncated domain, also
well matches the numerically computed $N(k)$ dependences, in spite of the
absence of any fitting parameter in the underlying interpolation [on the
contrary to the presence of parameter $\lambda $ in Eqs. (\ref{tanh}) and (%
\ref{tanh-1D}), which is used below in Fig. \ref{Fig3}]. A discrepancy in
the analytical and numerical shapes of the localized mode occurs at $\alpha
>1$ for bright solitons, although the wings are still well approximated by
the analytical ansatz in Fig. \ref{Fig1}(d), which is explained by its
compliance with the asymptotically exact expression (\ref{TFA}). In the
latter case, the discrepancy is naturally explained by the fact that the
smooth interpolation formula cannot accurately follow effects of the steep
modulation.

Direct simulations demonstrate that all the 1D LCW states found at $\alpha
\leq 1$, as well as their bright-soliton continuation at $\alpha >1$, are
completely stable, in accordance with the fact that they represent the
system's ground state. In addition, as seen in Fig. \ref{Fig1}(e), relations
$N(k)$ at different values of $\alpha $ obey the above-mentioned anti-VK
criterion, $dN/dk<0$, which provides the necessary stability condition for
the localized modes in the defocusing medium; it is actually a sufficient
one for ground states \cite{antiVK}.

Examples of 1D fundamental (single-notch) LDSs, with different values of the
modulation power $\alpha $, are displayed in Fig. \ref{Fig2}(a). It is seen
that the solitons' amplitude increases, the waist shrinks, and the decaying
tails sharpen with the increase of $\alpha $. The $N(k)$ curves for the
families of 1D fundamental LDSs, shown in Fig. \ref{Fig2}(d), along with
their TFA-predicted counterparts, given by Eq. (\ref{ND}),\ are obtained for
the truncated 1D system with $|x|<R=15$, covered by a grid of $1024$ points,
cf. Eq. (\ref{ND}). A (relatively small) mismatch between the numerical and
TFA curves is explained by the fact that the TFA ignores the norm defect
induced by the notch.

The comparison of typical numerically found profiles of the 1D fundamental
LDSs with their counterparts provided by the deformed-tanh interpolation (%
\ref{tanh}) is presented in Fig. \ref{Fig3}. It shows that the approximation
is very accurate, provided that the value of $\lambda $ in ansatz (\ref{tanh}%
) is selected as one which yields the best fit of the analytically predicted
LDS profile to the numerical one.

\begin{figure}[tbp]
\begin{center}
\includegraphics[width=1\columnwidth]{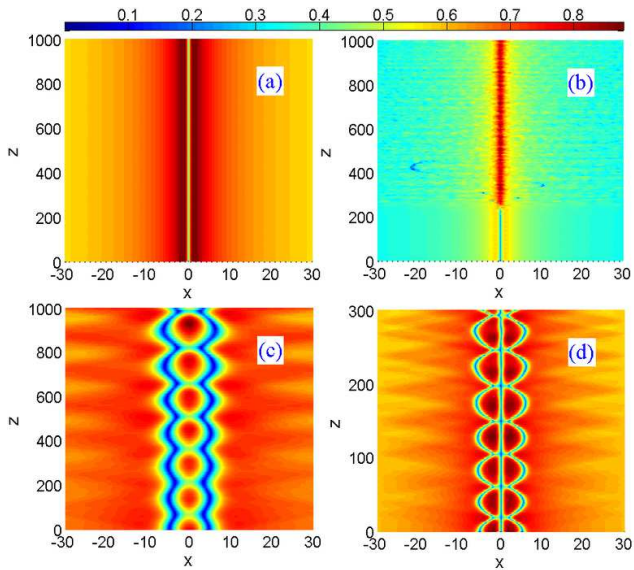}
\end{center}
\caption{(Color online) The evolution of stable and unstable 1D fundamental
LDSs is shown in (a) and (b), for $\protect\alpha =0.3$, $k=-1.9$, and $%
\protect\alpha =0.35$, $k=-2.4$, respectively. Panels (c) and (d) show
oscillations of weakly unstable perturbed 1D higher-order LDSs with
different numbers of notches: (c) two, for $\protect\alpha =0.25$, $k=-2.2$;
(d) three, for $\protect\alpha =0.3$, $k=-1$.}
\label{Fig5}
\end{figure}

Results for the stability analysis of the fundamental LDSs, collected in
Fig. \ref{Fig4}(a), demonstrate that they are stable at
\begin{equation}
\alpha <\alpha _{\max }^{\mathrm{(LDS)}}\approx 0.55,  \label{1D}
\end{equation}%
for values of $|k|$ which are not too large. Typical examples of the
evolution of stable and unstable 1D fundamental LDSs are displayed in Figs. %
\ref{Fig5}(a) and \ref{Fig5}(b), respectively. As can be seen in the latter
figure, the unstable fundamental LDS spontaneously loses its spatial
antisymmetry (i.e., the notch, at which $\left\vert \psi \left( x,z\right)
\right\vert $ was originally vanishing, gets filled), and quickly evolves
into an excited (oscillating) version of the LCW, i.e., \ a disturbed ground
state, with a maximum, rather than minimum, of $\left\vert \psi \left(
x,z\right) \right\vert ^{2}$, at $x=0$. 

It is relevant to stress that the conclusions concerning the stability of the 1D LDSs,
produced by the simulations performed in the domain of size $|x|\leq 30$, as shown 
in Fig. \ref{Fig5}, are virtually the same as obtained using a smaller domain
(not shown here in detail), $|x|\leq 10$,
(which is still essentially larger than the size of the corresponding LDSs), the shape of 
the respective stability area remaining the same in Fig. \ref{Fig4}(a). It is also relevant 
to stress that in those cases when the LDSs are unstable [Fig. \ref{Fig5}(c,d)],
the instability is clearly seen to set in the central segment of the integration 
domain, rather than arriving from the periphery. These findings confirm that conclusions 
about the stability of the LDSs in the infinite system may be based on the simulations 
performed in finite domains, in spite of the divergence of the total norm in the 
infinite system. The same conclusions are valid for the 2D setting. In particular, conclusions
concerning the instability boundary and development of 2D vortices (see Fig. \ref{Fig9} below)
are virtually the same for the domain with size $|x,y|\leq 20$ and for a smaller one (not shown 
in detail here) of size $|x,y|\leq 10$. In the latter case, the instability remains confined too
to the central core of the integration domain.
\begin{figure}[tbp]
\begin{center}
\includegraphics[width=1\columnwidth]{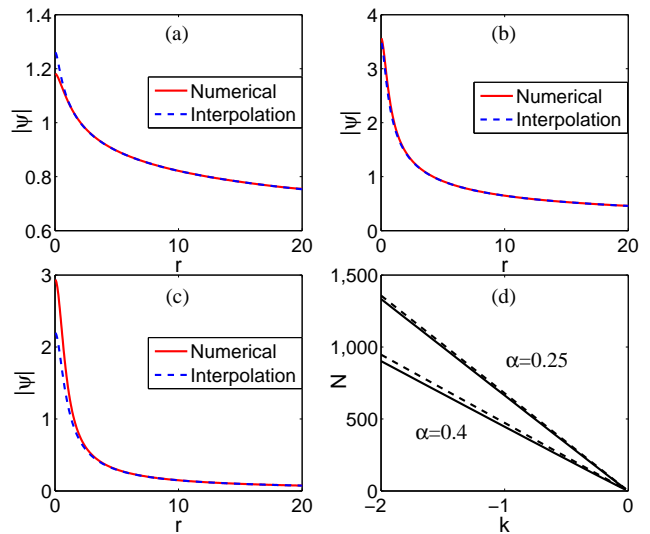}
\end{center}
\caption{(Color online) The comparison of the analytical interpolation,
given by Eq. (\protect\ref{LCW}), for 2D localized continuous-waves (LCW)
states, with their numerically found counterparts, for (a) $\protect\alpha %
=0.25$, $k=-1.2$; (b) $\protect\alpha =1$, $k=-4.2$; and (c) $\protect\alpha %
=2$, $k=-2.2$. (d) The numerically found dependence $N(k)$ for the 2D LCW
states in the truncated model, at different values of $\protect\alpha $.
Dashed lines display the analytical approximation for $N(k)$, as produced by
Eq. (\protect\ref{Nans1}).}
\label{Fig6}
\end{figure}

A noteworthy property of the 1D model is the existence of excited states in
the form of higher-order LDSs which are defined by the number of the
profile's zero-crossings (notches), starting from the single one for the
fundamental LDS. As is well known, the integrable NLS equation,
corresponding to $\alpha =0$, does not give rise to higher-order dark
solitons. Typical examples of the second- and third-order LDSs are displayed
in Figs. \ref{Fig2}(b) and \ref{Fig2}(c), respectively.

All the higher-order LDSs are unstable, but the instability may be weak. As
shown in Fig. \ref{Fig5}(c) and \ref{Fig5}(d), weakly unstable higher-order
dark solitons develop regular spatially symmetric oscillations, keeping the
initial number of notches. These plots clearly demonstrate that the
instability of the higher-order LDSs is explained by the interaction between
individual notches.

\begin{figure}[tbp]
\begin{center}
\includegraphics[width=1\columnwidth]{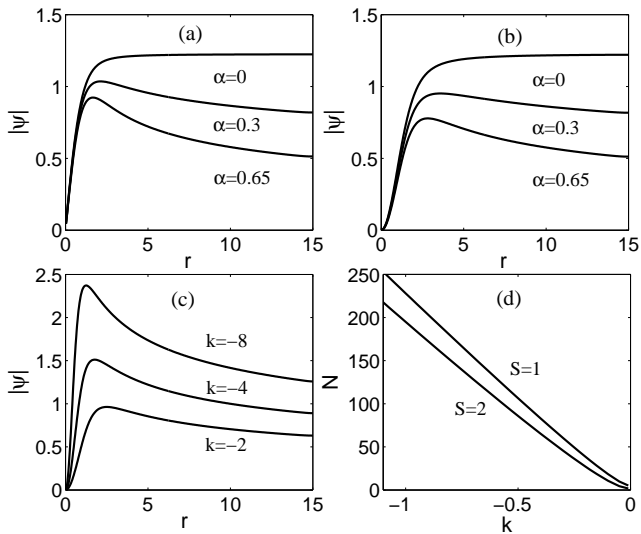}
\end{center}
\caption{Panels (a) and (b) display numerically found profiles of 2D
localized dark vortices (LDVs) with topological charges, severally, $S=1$
and $2$, propagation constant $k=-1.5$, and different values of $\protect%
\alpha $. The profiles of the usual vortices in the uniform space,
corresponding to $\protect\alpha =0$, are shown for comparison. (c) Profiles
of vortices with different values of $k$, for $\protect\alpha =0.6$ and $S=2$%
. (d) The norm of the vortex in the truncated system, with $r\leq R=15$, vs.
$k$, for the LDVs with $S=1,2$ and $\protect\alpha =0.4$. }
\label{Fig7}
\end{figure}

\section{Numerical results for two-dimensional localized continuous-wave and
dark-vortex modes}

In this section we focus on 2D LCW states and 2D LDVs, the latter
representing the most interesting modes in the 2D geometry. Noteworthy
results are obtained for the modes of both types.

The comparison of the analytical interpolation for the LCW, based on Eqs. (%
\ref{LCW}) and (\ref{r0}), with the numerically found solutions is shown in
Fig. \ref{Fig6}(a,b,c). Outer segments of the analytical shapes completely
overlap with their numerical counterparts, while the inner ones show minor
discrepancies at smaller and larger values of $\alpha $. Note that, similar
to the 1D LCWs [see Fig. \ref{Fig1}(a-c)], good accuracy is provided by the
analytical interpolation without the use of any fitting parameter. At $%
\alpha >2$, the 2D LCW state carries over into the fundamental bright
soliton, which was found in Ref. \cite{soliton-Defo1}. Similar to those 1D
counterparts, the 2D LCW states are found to be completely stable in their
entire existence region, $\alpha \leq 2$, which is easily explained by the
fact that they are ground states of the 2D setting.

\begin{figure}[tbp]
\begin{center}
\includegraphics[width=1\columnwidth]{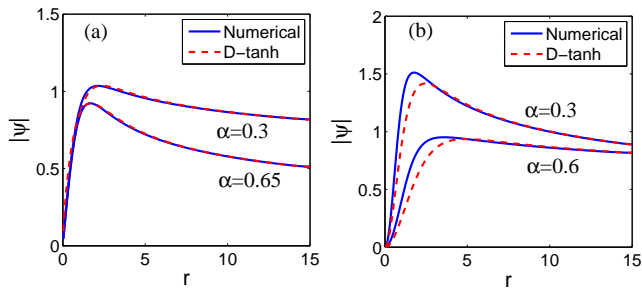}
\end{center}
\caption{(Color online) The comparison of numerically found profiles of the
2D LDVs with the fit provided by the deformed-tanh interpolation (\protect
\ref{tanh}) (labeled by \textquotedblleft D-tanh"): (a) $S=1$, $\protect%
\lambda =1.1$; (b) $S=2$, $\protect\lambda =1.5$. The propagation constant
is $k=-4$ for $\protect\alpha =0.6$ in (b), while $k=-1.5$ for the other
vortices shown here.}
\label{Fig8}
\end{figure}

Examples of LDVs with vorticities $S=1$ and $2$ are plotted, respectively,
in Figs. \ref{Fig7}(a) and (b). Both subplots include the standard dark
vortex in the uniform space, with $\alpha =0$ \cite{vortex,TFA}, for the
sake of comparison. It is seen that, with the increase of $\alpha $, the
shape of the weakly localized vortex sharpens, for both $S=1$ and $2$,
similar to the trend featured by the 1D LDS in Fig. \ref{Fig2}(a). On the
other hand, the amplitude of the vortex decreases, while in Fig. \ref{Fig2}%
(a) it was increasing with the growth of $\alpha $. The deformed-tanh
interpolation (\ref{tanh}) may be accurately fitted to the numerically found
shape of the LDVs with $S=1,$ as shown in Fig. \ref{Fig8}(a).

Results of the stability tests for perturbed LDVs with $S=1$ are collected
in Fig. \ref{Fig4}(b). The vortices are stable at $\alpha <\alpha _{\max
}^{(S=1)}\approx 0.90$ [cf. Eq. (\ref{1D})], for values of $|k|$ which are
not too large. A stability area was also found for the LDVs with $S=2$ (in
particular, it is bounded by $\alpha <\alpha _{\max }^{(S=2)}\approx 0.88$),
but its exact delineation requires excessively heavy simulations.

\begin{figure}[tbp]
\begin{center}
\includegraphics[width=1.025\columnwidth]{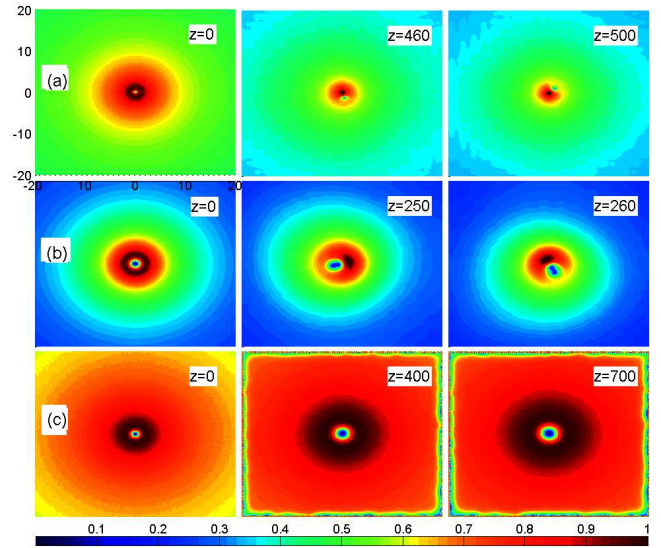}
\end{center}
\caption{(Color online) Snapshots of the perturbed evolution of unstable 2D LDVs
with different topological charges, $S$: (a) a vortex with $S=1,$ $\protect%
\alpha =0.51,$ $k=-2.7$ turns into the eccentric vortex orbiting the center;
(b) an unstable double vortex, with $S=2$, $\protect\alpha =1.2$, $k=-1.8$,
splits into a rotating pair of unitary vortices; (c) the transformation of
an unstable double vortex, with $S=2,$ $\protect\alpha =0.45,$ $k=-2.5$,
into a broader, apparently stable, double vortex. All the subplots are produced
in the domain size $|x,y| < 20$, as marked in panel (a).}
\label{Fig9}
\end{figure}

Different unstable vortices feature different scenarios of the instability
development. In Fig. \ref{Fig9}(a), an unstable LDV with $S=1$ transforms
into an eccentric vortex, with the pivot moving along a circular trajectory.
Furthermore, 
in Fig. \ref{Fig9}(b) an unstable double vortex ($S=2$) splits into a
rotating pair of unitary ones, which is typical for multiple vortices in
self-defocusing media \cite{vortex}. Lastly, in Fig. \ref{Fig9}(c), the
perturbed double vortex (again, the one with $S=2$) spontaneously broadens
and turns into a different, apparently stable, double LDV. The latter
scenario was found to be typical for the evolution of the LDVs which are
unstable because their $|k|$ is too large. 

\section{Conclusion}

It has recently been demonstrated that, opposite to the common belief, the
pure defocusing nonlinearity can support various robust bright solitons in
the space of dimension $D$, provided that the local strength of the
nonlinearity increases from the center to the periphery faster than $r^{D}$.
In previous works, only bright-soliton modes were theoretically investigated
in settings of this type. In this work, we have addressed loosely localized
modes, with the divergent norm, hence they may be categorized as LCWs, LDSs,
and LDVs (localized continuous waves, localized dark solitons, and localized
dark vortices, respectively). Such modes are supported by the local
nonlinearity strength growing slower than, or exactly as, $r^{D}$. The
corresponding nonlinearity landscapes may be realized for light waves in
nonlinear photonic crystals or in inhomogeneously doped optical media, and
for matter waves in BEC, controlled by means of the spatially inhomogeneous
Feshbach resonance. The LCWs, which represent the ground state in the 1D and
2D geometries, LDSs (both fundamental and higher-order ones, characterized
by multiple notches in the 1D case), and LDVs, with vorticities $S=1$ and $%
S=2$, have been constructed by means of analytical approximations and
numerical methods. The relevant analytical methods are the TFA\
(Thomas-Fermi approximation) for universal tails of all modes, and the
interpolation between analytical approximations available far from and close
to the center, for the LCW states in 1D and 2D, as well as for 1D LDS and 2D
LDV. In particular, the interpolation based on the deformed-tanh expression
provides for a very accurate fit to numerically found 1D LDS and 2D LDV
profiles (the latter one fits well for $S=1$). Stability areas for these
modes, and instability development scenarios for unstable ones, have been
identified by means of systematic simulations of the perturbed evolution. In
particular, the LCWs are completely stable in 1D and 2D alike, which is
readily explained by the fact that they serve as ground states in the
respective settings.

In terms of BEC, the present analysis may be extended for the 3D setting,
with $\alpha <3$ in Eq. (\ref{asympt}). On the other hand, it may be
interesting to extend the analysis to loosely localized states in 1D and 2D
models with spatially modulated nonlocal nonlinearity, where, in particular,
the stabilization of vortex solitons is also a relevant problem \cite%
{nonlocal}.

\section{Acknowledgment}

We appreciate valuable discussions with G. Assanto. The work of J. Z. is
supported by NSFC, China (project Nos. 61690222,61690224,11204151), by the Initiative
Scientific Research Program of the State Key Laboratory of Transient Optics
and Photonics, and partially by the Youth Innovation Promotion Association
of the Chinese Academy of Sciences (project No. 2016357) and the CAS/SAFEA
International Partnership Program for Creative Research Teams. The work of B. A. M.
is supported, in part, by grant No. 2015616 from the joint program in
physics between the National Science Foundation (US) and Binational
(US-Israel) Science Foundation.

\end{document}